# Cryptographic Application of Elliptic Curve with High Rank

Xiaogang CHENG[1], Ren GUO[2], Zuxi Chen[1]

1. College of Computer Science and Technology, Huaqiao University,
Xiamen 361021, China

2. College of Business Administration, Huaqiao University, Quanzhou 362021, China

**Abstract:** Elliptic curve cryptography is better than traditional cryptography based on RSA and discrete logarithm of finite field in terms of efficiency and security. In this paper, we show how to exploit elliptic curve with high rank, which has not been used in cryptography before, to construct cryptographic schemes. Concretely we demonstrate how to construct public key signature scheme with hierarchy revocation based on elliptic curve with high rank, where the rank determines the height of the revocation tree. Although our construction is not very efficient in some sense, our construction shows elliptic curve with high rank is valuable and important for cryptographic usage. The technique and assumption presented can surely be used for other cryptographic constructions.

## 1. Introduction

At the beginning of cryptographic applications of EC (Elliptic Curve), it was used to factor large integers [1], which can be used to break the security of RSA. Later elliptic curve was introduced into cryptography in a constructive way by Koblitz and Miller in about 1986 [2][3], to replace the traditional commonly used discrete logarithm problem in finite field by the same problem in elliptic curve, which is safer and more efficient.

Today, there are many research works and practical applications of elliptic curve in many areas, such as vehicular sensor network [4], Internet of Things and healthcare [5], image encryption [6], blockchain [7], wireless sensor networks [8], smart cards [9], smart home [10] etc. In 2001, another important technique of elliptic curve, namely bilinear paring, was introduced into cryptography by Boneh et. al. to build Identity-Based Encryption [11] and short signatures [12]. And nowadays bilinear paring also has many applications in many areas, such as clouding computing [13], vehicular ad-hoc network [14], succinct non-interactive argument of knowledge [15,16, 17], healthcare [18], blockchain and cloud storage [19].

There are also some standardizations on elliptic curve, like FIPS 186-5 [20,21] on digital signature released by NIST (National Institute of Standards and Technology). And NIST standardization on bilinear paring is also underway [22]. Chinese encryption administration released SM2 elliptic curve algorithm in 2010 [23], and become a Chinese national standard in 2016 [24].

Despite many properties and techniques of elliptic curve have been used in cryptography as mentioned above. One important aspect of elliptic curve, namely rank, has not been exploited for cryptographic applications. It is well-known that the group of rational points of elliptic curve is isomorphic to $G_{TOR} \otimes Z^r$, where $G_{TOR}$ is the torsion group of finite order, and $r$ is the rank. And it is unknown whether $r$ can be arbitrary large [25], and there is a very famous conjecture, namely BSD conjecture [25], about the rank and the L-function. In this paper, we exploit elliptic curve with high rank for cryptographic usage. Namely we construct a PKS (Public Key Signature) scheme with hierarchical revocation mechanism based elliptic curve with high rank. Of course, this is just the beginning of exploiting the rank of elliptic curve, we believe that the techniques used in this paper can certainly be used in many other cryptographic applications.

Some preliminaries are given in section 2. Our concrete construction of PKS with hierarchical revocation is presented in section 3. In section 4, the security of our scheme is analyzed and proved. And we conclude in section 5.

**2. Preliminaries**

**Definition 1.** PKS with hierarchical revocation.

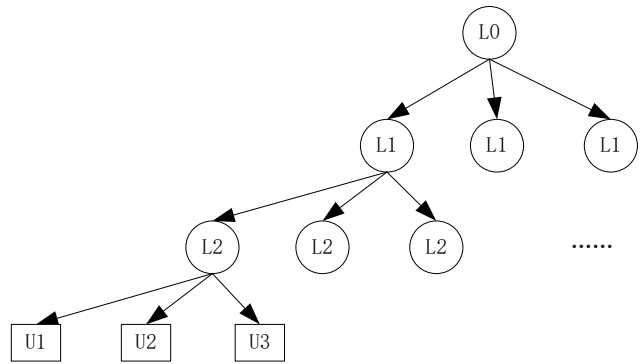

Fig. 1. A simple hierarchical structure for an organization

There is a GM (Group Manager) who is responsible for generating PK/SK pairs for each member of an organization with hierarchical structure as shown in Fig. 1. With its secret key gotten form the GM, each member can generate legitimate signature. Later GM can carry out the following different kinds of revocation:

1. Revocation of one member. The GM can revoke the signing ability of a member. After which, the revoked member can no longer generate legitimate signature.

2. Revocation of a group. The GM can also revoke a group in the hierarchical

structure. After which all the members of the group will lose their signing ability. For example, GM can revoke the left most group of L2 in Fig. 1. Then all the members of group, i.e. U1, U2 and U3 will be revoked. And GM can also choose to revoke a larger group, i.e. groups in L1, then all the members belong to this larger group will be revoked. This will be much more efficient than revoke all the members individually by using the first kind of revocation.

Let's consider the following realistic scenarios. A large corporation will have many companies, and each company will have many departments, such as the sales department, HR department, finance department and IT department etc. And each department will have some employees. When an employee joins the corporation, he or she will get his/her PK/SK from the GM of the corporation. Then later when one employee quit his job, the GM can revoke his signing ability by using the first kind of revocation above. When a company decide to cut off a whole department, the GM of the corporation can revoke all the members of the department by using the second kind of revocation above. Similarly, the GM can also revoke the signing ability of all the employees of a whole company, if the corporation decide to close the company.

**Security of PKS with hierarchical revocation:**

1) Non-forgeablility: Corresponding any PK that is signed by the GM, only the owner of the corresponding SK can generate legitimate signature. Noone else can generate legitimate signature corresponding the PK

2) Revocation security:

- After revocation of a member, the revoked member will lose the signing ability. I.e. he will not be able to generate legitimate signature for any message.
- After revocation of a group, all the members of the group will lose their signing ability. Each member will not be able to generate legitimate signature for any message.

**EC with high rank**

It is well-known that the rational points of an elliptic curve form a group which is isomorphic to $G_{TOR} \otimes Z^r$, where r is the rank of the elliptic curve. For example the following EC is a curve with rank 0, i.e. there are only finite many rational points:

$$y^2 = x^3 + 3x$$

The following are examples of elliptic curve with rank 1 and 2:

$$y^2 = x^3 + 877x, \quad rank\ 1$$

$$y^2 = x^3 + 73x, \quad rank\ 2$$

Currently it is unknown whether rank can be arbitrary large. Also, it is very hard to find elliptic curve with very high rank. The following is an example of EC with rank 14:

$$y^2 = x^3 + 402599773876907010169104272724483x$$

The following EC is of rank 28 at least [26]:

$$y2 + xy + y = x3 - x2 + bx + c$$

$b = -20067762415575526585033208209338542750930230312178956502,$

$c = 34481611795030556467032985690390720374855944359319180361266008296291939448732243429$

Current highest record is an EC with rank 29 at least [27,28], constructed by Elkies and Klagsbrun in 2024.

## 3. Our Construction of PKS with hierarchical revocation

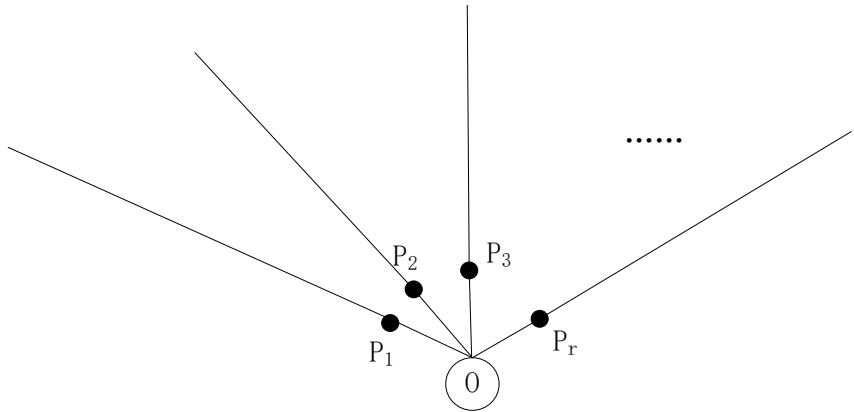

Fig. 2. Schematic diagram for elliptic curve with rank r and generators

**1.Setup**

1) GM choose an EC with rank $r$ (see Fig. 2), where $r$ will determine the height of the revocation tree, and the r generators of infinite order are:

$$P_1, P_2, P_3, \dots, P_r$$

2) GM choose a large random prime number $p$, modulate which it is hard to compute non-trivial relations among the r generators. (see Assumption 1 in next section)

3) For each top-level department, GM choose a random co-dimension 1 hyperplane:

$$a_0 + a_1 x_1 + a_x x_2 + \cdots + a_r x_r = 0$$

Later, the secret keys of all the members of a top-level department will lie on the corresponding hyperplane. For each second level department, GM will choose a random co-dimension 2 hyperplane on the hyperplane of the corresponding top-level department. For example, the intersection of the following 2 hyperplane:

$$a_0 + a_1 x_1 + a_2 x_2 + \cdots + a_r x_r = 0$$

$$b_0 + b_1 x_1 + b_2 x_2 + \cdots + b_r x_r = 0$$

And similarly for the 3rd, 4th level, until the rth level department, GM generate a random co-dimension k hyperplane for each kth department, which lie on the co-dimension $k-1$ hyperplane of the father node on the hierarchical tree.

2. Join

When a new member joins the organization, the GM will generate a PK/SK pair for the new member. The SK is just a tuple of $r$ random integers $(c_1, c_2, \ldots, c_r)$, which will lie on the co-dimension r-1 hyperplane of the $rth$ department that the member belongs to. PK is just:

$$PK = c_1 P_1 + c_2 P_2 + c_3 P_3 + \cdots + c_r P_r$$

Then GM publish the PK, and sign it to certify its validity.

3. Sign

The ordinary signature for a message $m$ is just the following zero-knowledge proof of knowledge:

$$ZK\{(c_1, c_2, c_3, \ldots, c_r): PK = c_1 P_1 + c_2 P_2 + c_3 P_3 + \cdots + c_r P_r\}(m)$$

4. Revocation

1) To revoke a member, the GM can just put the public key of the revoked member to the revocation list. Then later any verifier can refute a signature if he finds that the verification PK is in the RL.

2) To revoke a group.

As mentioned above, when generating SK/PK pairs, the GM deliberately choose the SKs of the same department to lie on the same hyperplane. I.e., there is a secret random co-dimension k hyperplane for each department of level k. For example, the level-1 department has a co-dimension 1 (namely dimension r-1) hyperplane with equation:

$$a_0 + a_1 x_1 + a_x x_2 + \cdots + a_r x_r = 0$$

Then this level-1 department can be revoked by putting the hyperplane to the RL, and ask all the members other than this revoked department to prove that his secret key does not satisfy this equation in later signatures, i.e.:

$$ZK\{(x_1, x_2, x_3, \ldots, x_r): PK = x_1P_1 + x_2P_2 + x_3P_3 + \cdots + x_rP_n$$

$$\wedge\ a_0 + a_1x_1 + a_xx_2 + \cdots + a_rx_r \neq 0\}(m)$$

Namely, the signer has to prove that his knows the secret key corresponding to the PK, also the secret keys has not been revoked (i.e., his secret key is not on the revoked hyperplane).

To be more concrete, now suppose we have three layers as shown in Fig. 1. GM is in L0. For each L1 group, GM can use a random co-dimension 1 hyperplane. In this case, r=3, it is a 2-dimension plane. Of course, on this plane there are many lines (co-dimension 2 hyperplane), then for each L2 group, GM can choose a random line on the corresponding L1 plane.

I.e., for each Level 1 (L1) group, GM choose a random plane and random prime integer $q$ (see next section for the selection of $q$):

$$Ax_1 + Bx_2 + Cx_3 + D = 0\ mod\ q$$

Then, later all the member of the L1 group (including subgroups) will have secret keys $(c_1, c_2, c_3)$ lying on this plane.

For each L2 group of a L1 group, GM randomly choose a random line on the corresponding plane.

$$Ax_1 + Bx_2 + Cx_3 + D = 0\ mod\ q \qquad (1)$$

$$A'x_1 + B'x_2 + C'x_3 + D' = 0\ mod\ q \qquad (2)$$

Hence GM will generate secret keys of this L2 group (namely users U1, U2 and U3) to lie on this line (also on the L1 plane above, of course)

Revoke one person: The GM just put the public key of the revoked member to the RL. Then later when someone make a signature using the revoked PK, any verifier can find it out from the RL and refute the signature.

If all the member of one L2 group are to be revoked. It is tedious to put all the public keys of all the members to the RL. Instead, the GM can choose to revoke the whole group. To Revoke a L2 group, GM put the hyperplane of the L2 group to the RL. Then later signer should prove that his secret key is not on the revoked hyperplane, namely:

$$ZK\{(x_1, x_2, x_3): PK = x_1P_1 + x_2P_2 + x_3P_3\ \wedge$$

$$(Ax_1 + Bx_2 + Cx_3 + D \neq 0 \vee A'x_1 + B'x_2 + C'x_3 + D' \neq 0)\ \}(m)$$

If all the L2 groups belonging to a L1 group are revoked, then the hyperplanes corresponding to these L2 groups in RL can be replaced by just the hyperplane of the L1 group, i.e. $Ax_1 + Bx_2 + Cx_3 + D = 0$. Then later signer should prove that his secret key is not on this L1 hyperplane:

$$ZK\{(x_1, x_2, x_3): PK = x_1P_1 + x_2P_2 + x_3P_3 \wedge Ax_1 + Bx_2 + Cx_3 + D \neq 0 \}(m)$$

In general, RL is consisted of revoked public keys of individual member, and revoked hyperplane of a group. New signature should prove that the secret key used is not on any hyperplane in the RL:

$$ZK\{(x_1, x_2, \dots, x_r): PK = x_1P_1 + x_2P_2 + \cdots + x_rP_r \wedge$$

$$(x_1, x_2, \dots, x_r) \text{ is not on hyperplane } P \;\forall P \in RL \}(m)$$

**A Toy example**

It is known that the rank of the following rational elliptic curve is two:

$$y^2 = x^3 + 17$$

And over $Q$, there are no torsion point on this curve. The two generators are:

$$P_1 = (-2,3), \quad P_2 = (2,5)$$

The first few rational points generated by $P_1$ are:

Table 1. Rational points generated by $P_1$

|       | $nP_1$ |
|-------|--------|
| n=1   | $(-2,3)$ |
| n=2   | $(8,-23)$ |
| n=3   | $(\frac{19}{25}, \frac{522}{125})$ |
| n=4   | $(\frac{752}{529}, -\frac{54239}{12167})$ |
| n=5   | $(\frac{174598}{32761}, \frac{76943337}{5929741})$ |
| n=6   | $(-\frac{4471631}{3027600}, -\frac{19554357097}{5268024000})$ |
| n=7   | $(\frac{12870778678}{76545001}, \frac{1460185427995887}{669692213749})$ |

As we can see that the "height" of the points grows very fast. Choose a random prime $p = 3123456773$, mod this p we get:

Table 2. Points over $F_{3123456773}$ generated by $P_1$

|  | $nP_1 \bmod p$ |
|---|---|
| n=1 | (3123456771, 3) |
| n=2 | (8, 3123456750) |
| n=3 | (2748641961, 2148938264) |
| n=4 | (743961350, 253378136) |
| n=5 | (1176218259, 691053659) |
| n=6 | (2180670293, 2607412353) |
| n=7 | (128580328, 2472269909) |

Similarly for $P_2 = (2,5)$, we have:

Table 3. Rational and $F_{3123456773}$ points generated by $P_2$

| $nP_2$ | $nP_2 \bmod p$ |
|---|---|
| (2, 5) | (2, 5) |
| $(-\frac{64}{25}, \frac{59}{125})$ | (2248888874, 2923555540) |
| $(\frac{5023}{3249}, -\frac{842480}{185193})$ | (2602399966, 2884651714) |
| $(\frac{38194304}{87025}, -\frac{236046706033}{25672375})$ | (1188080486, 863393529) |
| $(\frac{279124379042}{111229587121}, \frac{212464088270704525}{37096290830311831})$ | (842290081, 2500317348) |
| $(-\frac{22792283822695031}{9224204064998400}, \frac{1225613646951190271274203}{885917648237503131648000})$ | (2145964735, 2073955284) |
| $(\frac{17206060394388022298882}{15290847667056681428641}, -\frac{8116122042886721305956245646487115}{18908076145393139649196885319125 61})$ | (759645483, 758431348) |

Now suppose a company has the following three departments, Financial Department, Human Resource Department and Department of Engineering.

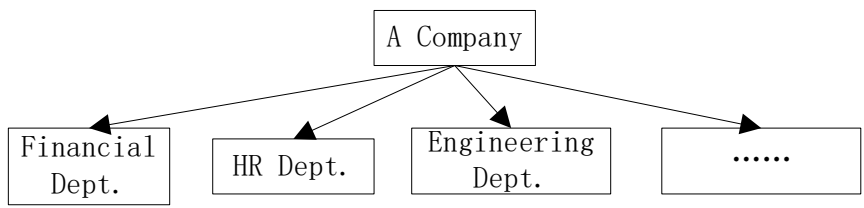

Fig. 2. A simple hierarchical structure for a company

For each department, a random hyperplane is assigned for generating PK/SK pairs for the stuff of this department. For example, the following three hyperplane can be assigned to the three departments of the company respectively (Note that in this simple toy example we make $p = q$ for simplicity, see next section for how to choosing $q$):

$$48x_1 + 79x_2 + 123 = 0 \bmod p$$

$$36x_1 + 139x_2 + 752 = 0 \bmod p$$

$$58x_1 + 32x_2 + 937 = 0 \bmod p$$

Suppose there are two new employees for the Financial Dpt., two random points on the first line can be generated easily as the secret keys for the two new employees:

$$SK_1 = (x_{11}, x_{21}) = (3257, 2774256590)$$

$$SK_2 = (x_{12}, x_{22}) = (6789, 118608156)$$

Since:

$$48 * 3257 + 79 * 3083917365 + 123 = 0 \bmod p, p = 3123456773$$

$$48 * 6789 + 79 * 118608156 + 123 = 0 \bmod p, p = 3123456773$$

The corresponding public keys are the points on the elliptic curve:

$$PK_1 = x_{11}P_1 + x_{21}P_2 = 3257P_1 + 2774256590P_2$$

$$PK_2 = x_{12}P_1 + x_{22}P_2 = 6789P_1 + 118608156P_2$$

I.e.,

$$PK_1 = (1385928692, 2187054458)$$

$$PK_2 = (2132129612, 2902520269)$$

It is easy to revoke a member, the GM just put the public key of the revoked member to the RL, for example put $PK_1 = (1385928692, 2187054458)$ to the RL to revoke the first employee above. Then later verifier of a signature can easily check if the corresponding public key is in the RL or not. If so, the signature can easily be refuted by the verifier.

To revoke the financial department, the GM put the hyperplane corresponding to this department, namely:

$$48x_1 + 79x_2 + 123 = 0 \bmod p$$

to the RL. And then later signature should prove that the secret key used is not on this hyperplane, namely:

$$ZK\{(x1, x2): x_1 P_1 + x_2 P_2 = PK_1 \land 48x_1 + 79x_2 + 123 \neq 0 \bmod p\}$$

Note that after this department revocation, all the stuff of the financial department will lose their signing ability. Since all their secret keys are lying on the revoked hyperplane, they will not be able to generate a zero-knowledge above.

## 4. Security Proof

**Assumption 0**: Modulate a large random prime number, it is hard to calculate non-trivial linear relation among the generators of an EC with high rank. I.e., for an EC with rank r and generators $P_1, P_2, \ldots, P_r$, it is hard to compute integers $x_1, x_2, \ldots, x_r$ not all zero, such that:

$$x_1 P_1 + x_2 P_2 + \cdots + x_r P_r = 0, (x_1, x_2, \ldots, x_r) \neq (0,0,\ldots,0)$$

Based on this assumption, we can prove the security of our PKS with hierarchical revocation.

The problem of the above assumption is that the order of a generator $P_i$ might be easy to figure out. Since it must divide the order of the finite group of EC over finite field F_P. If so, we may have an easy equation like:

$$0P_1 + 0P_2 + \cdots + Order(P_i) * P_i + \cdots + 0P_r = 0$$

which violate assumption 0. To circumvent this loophole, we may impose further restrictions on the magnitude of $(x_1, x_2, \ldots, x_r)$. Namely, we restrict that each $x_i$ should be less than a prime number $q$, which is significantly less that the possible order of each generator $P_i$. Hence:

**Assumption 1**: In assumption 0, we further restrict that each $x_i$ is in the range $[0, q)$, where q is significantly less then the order of any $P_i$, then it is hard to compute the same $(x_1, x_2, \ldots, x_r)$ as assumption 0 and at least two $x_i$ are not zero. (For simplicity, we omit this minor technical detail in later proof and narrative.)

**Theorem 1**. The PKS scheme with hierarchical revocation constructed above satisfy non-forgeability based on the Assumption 1.

**Proof:** Recall that the signature is just the following zero-knowledge proof of knowledge:

$$ZK\{(x_1, x_2, x_3, \ldots, x_n): PK = x_1 P_1 + x_2 P_2 + x_3 P_3 + \cdots + x_n P_n\}(m)$$

Based on the security properties of ZK (completeness and soundness). If a forged signature passed the verification, then there must be a legitimate witness which is known to adversary, and different with the original witness known to the signer, i.e. there is a

tuple $(x'_1, x'_2, x'_3, \ldots, x'_n) \neq (x_1, x_2, x_3, \ldots, x_n)$ satisfy the same equation:

$$PK = x'_1 P_1 + x'_2 P_2 + x'_3 P_3 + \cdots + x'_n P_n$$

Since $(x'_1, x'_2, x'_3, \ldots, x'_n) \neq (x_1, x_2, x_3, \ldots, x_n)$, at least one term of the following tuple is not zero:

$$(x_1 - x'_1, x_2 - x'_2, x_3 - x'_3, \ldots, x_n - x'_n)$$

Without loss of generality, we suppose $x_1 - x'_1 \neq 0$, then:

$$P_1 = \frac{-(x_2 - x'_2)P_2 - (x_3 - x'_3)P_3 - \cdots - (x_n - x'_n)P_n}{(x_1 - x'_1)}$$

For example, if $n = 2$, then discrete logarithm between $P_1$ and $P_2$ can be computed. Since we have:

$$(x_1 - x'_1)P_1 + (x_2 - x'_2)P_2 = 0$$

With this equation, it is simple to get:

$$P_1 = \frac{-(x_2 - x'_2)}{(x_1 - x'_1)} P_2$$

This clearly violate our assumption above which says that it is hard to compute nontrivial linear relations among the generators of each branch. □

**Theorem 2**. The PKS scheme with hierarchical revocation constructed above satisfy revocation security based on the Assumption 1. Namely revocation members or members of a revoked group cannot generate legitimate signature.

**Proof:** 1) Revocation security of one member is easy to prove. The GM just put the public key of the revoked member to the RL (Revocation List). After that, it is easy for any verifier to refute any signature whose verification key is in the RL.

2) Group revocation security. For group revocation, the GM put the hyperplane for the revoked group to the RL. And later all the legitimate signer should prove that his secret key is not on the revoked hyperplane in the RL (we use $f(\vec{x})$ to represent a hyperplance).

$$ZK\{(x_1, x_2, \ldots, x_n): PK = x_1 P_1 + x_2 P_2 + \cdots + x_n P_n \wedge f(\vec{x}) \neq 0, \forall f(\vec{x}) \in RL\}(m)$$

For this zero-knowledge proof to pass the verification, the revoked signer has to know a different $(x'_1, x'_2, x'_3, \ldots, x'_n) \neq (x_1, x_2, x_3, \ldots, x_n)$, since the original witness (x1,x2,..,xn) is on the hyperplane $f(\vec{x})$, which satisfy:

$$PK = x'_1 P_1 + x'_2 P_2 + x'_3 P_3 + \cdots + x'_n P_n$$

$$f(\vec{x'}) \neq 0, \forall f(\vec{x}) \in RL$$

This is harder than proof of number 1. Since this time that adversary not only have to compute $\vec{x'}$ which is not equal to $\vec{x}$, but also he has to make sure that the new $\vec{x'}$ is not on any revoked hyperplane. So simply by the first equation, a not-trivial linear relation can be gotten, this clearly violate our assumption above. □

## 5. Conclusions

In this paper we exploit yet another technique of EC for the cryptographic applications, namely the rank which has not been used in cryptography before. Concretely we present a mathematical assumption which says that it is hard to calculate non-trivial relations among the independent generators of an EC with high rank. Then based on this assumption, we present a PKS with hierarchical revocation.

We believe that our construction based on EC with high rank is just a beginning, the technique and assumption presented in this paper can surely be used in many other scenarios and applications in cryptography. And cryptographic usage of EC with high rank will also provide an impetus for searching for EC with higher and higher rank, which is a hard problem in mathematics.

Our construction of PKS with hierarchical revocation is not very efficient, since we make use of the general theorem of zero knowledge proof system for NP language. In the future, we plan to improve the efficiency by making use of efficient zero knowledge proof system, such as SNARK etc. Then our construction can be used in practice. Another further improvement is to use polynomial with high degree, instead of just linear equations (i.e. hyperplanes), to support more revocation functionalities. For example, two high degree polynomial can have multiple intersections points, then each intersection point can represent an employee who belong to multiple departments. Further studies on the security of our assumption, concrete curve which satisfy our assumption and ready for practical usage are also worth pursuing.